\documentclass{JINST}

\RequirePackage{xspace}

\def\gev    {\ensuremath{\mathrm{\,Ge\kern -0.1em V}}\xspace}
\def\mev    {\ensuremath{\mathrm{\,Me\kern -0.1em V}}\xspace}
\def\xrad   {\ensuremath{\,X_0}\xspace}
\def\lint   {\ensuremath{\,\lambda}\xspace}
\def\ma     {\ensuremath{{\rm \,m}^2}\xspace}
\def\cma    {\ensuremath{{\rm \,cm}^2}\xspace}
\def\mum    {\ensuremath{{\rm \,{\mu}m}}\xspace}
\def\cm     {\ensuremath{{\rm \,cm}}\xspace}
\def\mm     {\ensuremath{{\rm \,mm}}\xspace}

\title{Monte carlo study of the physics performance of a digital hadronic 
  calorimeter}

\author{Catherine Adloff, Jan Blaha\thanks{Corresponding author.}, 
Jean-Jaques Blaising, Maximilien Chefdeville, Ambroise Espargili\`{e}re, 
Yannis Karyotakis\\
\llap{$ $}Laboratoire d'Annecy-le-Vieux de Physique des Particules, 
Universit\'{e} de Savoie, CNRS/IN2P3,\\
9 Chemin de Bellevue 74980 Annecy-le-Vieux, France\\
E-mail: \email{jan.blaha@lapp.in2p3.fr}}

\abstract{A digital hadronic calorimeter using MICROMEGAS as active 
elements is a very promising choice for particle physics experiments at 
future lepton colliders. These experiments will be optimized for application 
of the particle flow algorithm and therefore require calorimeters with very 
fine lateral segmentation. A 1\ma prototype based on MICROMEGAS chambers 
with 1$\times$1\cma readout pads is currently being developed at LAPP. The 
GEANT4 simulation of the physics performance of a MICROMEGAS calorimeter is 
presented. The main characteristics, such as energy resolution, linearity and 
shower profile, have been carefully examined for various passive materials 
with pions over a wide energy range from 3 to 200\gev. The emphasis is put on 
the comparison of the analog and digital readout.}


\begin{document}

\section{Introduction}
Future particle physics experiments at the International Linear Collider 
(ILC)~\cite{ilc} will employ the Particle Flow Algorithm (PFA) to reach
a jet energy resolution of $30\%/\sqrt{E}$. In order to achieve an optimal PFA 
performance, a highly granular hadronic calorimeter with a good shower 
separation is required. One of the suitable and affordable choice for an 
active part of the hadronic calorimeter is a thin gaseous detector with 
embedded {\it digital} (1-bit) or {\it semi-digital} (2-bit) readout. This 
concept allows the construction of the so-called Digital Hardronic CALorimeter 
(DHCAL) with very fine granularity (a cell size of about 1\cma) providing 
high MIP efficiency, low hit multiplicity as well as negligible performance 
degradation due to high dose rates, hadronic showers and aging. 

One of the promising candidate for a DHCAL is the MICRO MEsh GAseous Structure
(MICROMEGAS) which is a micro-pattern gaseous detector~\cite{micromegas}. 
Prototypes with 1$\times$1\cma anode pads, currently under development at LAPP, 
consist of a commercially available 20\mum thin woven stainless steel mesh 
which separates the 3\mm drift gap from the 128\mum amplification gap filled 
by an Argon/Isobutane (95/5) gas mixture. The readout electronics is embedded 
on the PCB below the anode, and thus creates a compact detector of 8\mm 
thickness. The sampling calorimeter equipped with such a detector is used for 
this study. This allows the first qualitative view on DHCAL global performances.

\section{Calorimeter geometry and simulation tools}
The geometry of the hadronic calorimeter, originally proposed for the SiD 
detector~\cite{sid}, was adapted for this study with various absorber 
materials (Fe, W, and Pb). The depth of the calorimeter, which is in SiD
design 4.5\lint (40 absorber plates) was extended up to 9\lint 
(80 absorber plates) and hence the results obtained may also serve for the 
CLIC detector~\cite{clic} which operates at a higher center-of-mass energy. 
As an active medium the MICROMEGAS detector has been chosen and the prototype 
geometry described above was implemented in the simulation. The absorber 
thickness in terms of interaction length \lint is equal for all three 
absorbers, and therefore the total length of the calorimeter varies depending 
on the passive material. In case of Fe absorber, where two 2\mm thick steel 
cover plates of the MICROMEGAS chamber are supposed to be a part of the passive 
layer, the total calorimeter length is 200\cm. For W and Pb absorbers the steel 
covers were replaced by aluminum ones and thus the total length is about 
170\cm and 239\cm, respectively. A lateral size of 200$\times$200\cma is 
equal for all three calorimeters.    
    
Monte Carlo data for negative pions in a wide energy range from 3 to 200\gev 
were generated by a GEANT4-based simulator SLIC with LHEP physics list. The 
generated data, around 20,000 events per energy for each calorimeter 
configuration, were subsequently reconstructed and analyzed using the 
org.lcsim framework~\cite{slic}. Since the conversion from energy deposited 
in 3\mm gas gap to charge and electronics digitization were not 
included in the simulation, the so called {\it analog} and  {\it digital} 
readout represent the deposited energy (in\mev) in gas gap or the number of 
counted hits\footnote{One hit is counted only when the deposited energy in a 
cell is higher than a given threshold} in 1$\times$1\cma cells, respectively. 
Both quantities are considered only when a readout threshold of 0.1 MIP MPV is 
reached.  

\section{Energy shower profile}
\subsection{Longitudinal and lateral shower profile}
Longitudinal and lateral energy shower profiles were studied with various 
absorber materials in a wide energy range for both readouts. 

\begin{center}              
  \begin{figure}[htb]
    \includegraphics[width=0.49\columnwidth]{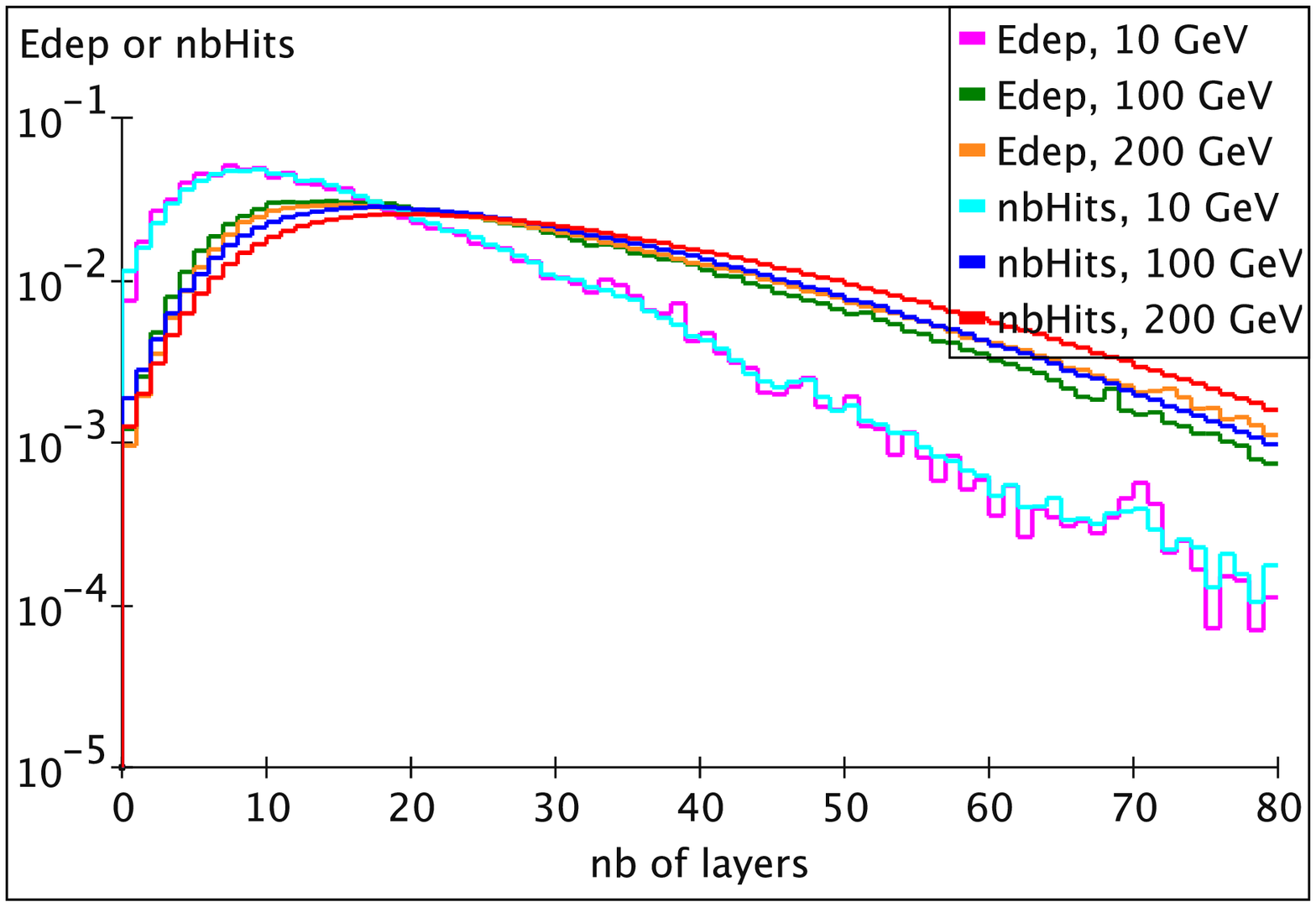}
    \hfill
    \includegraphics[width=0.49\columnwidth]{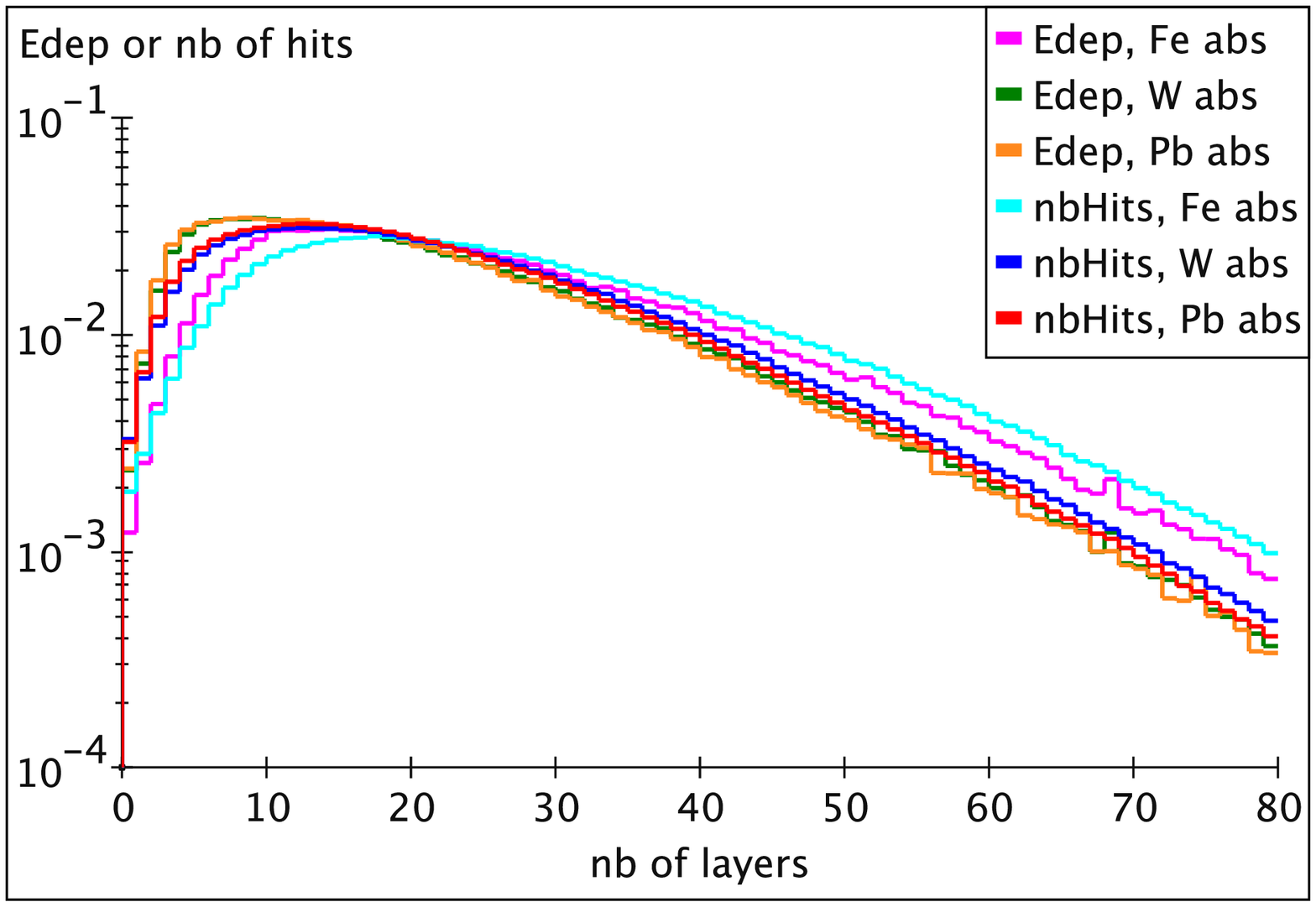}\\ 
    \includegraphics[width=0.49\columnwidth]{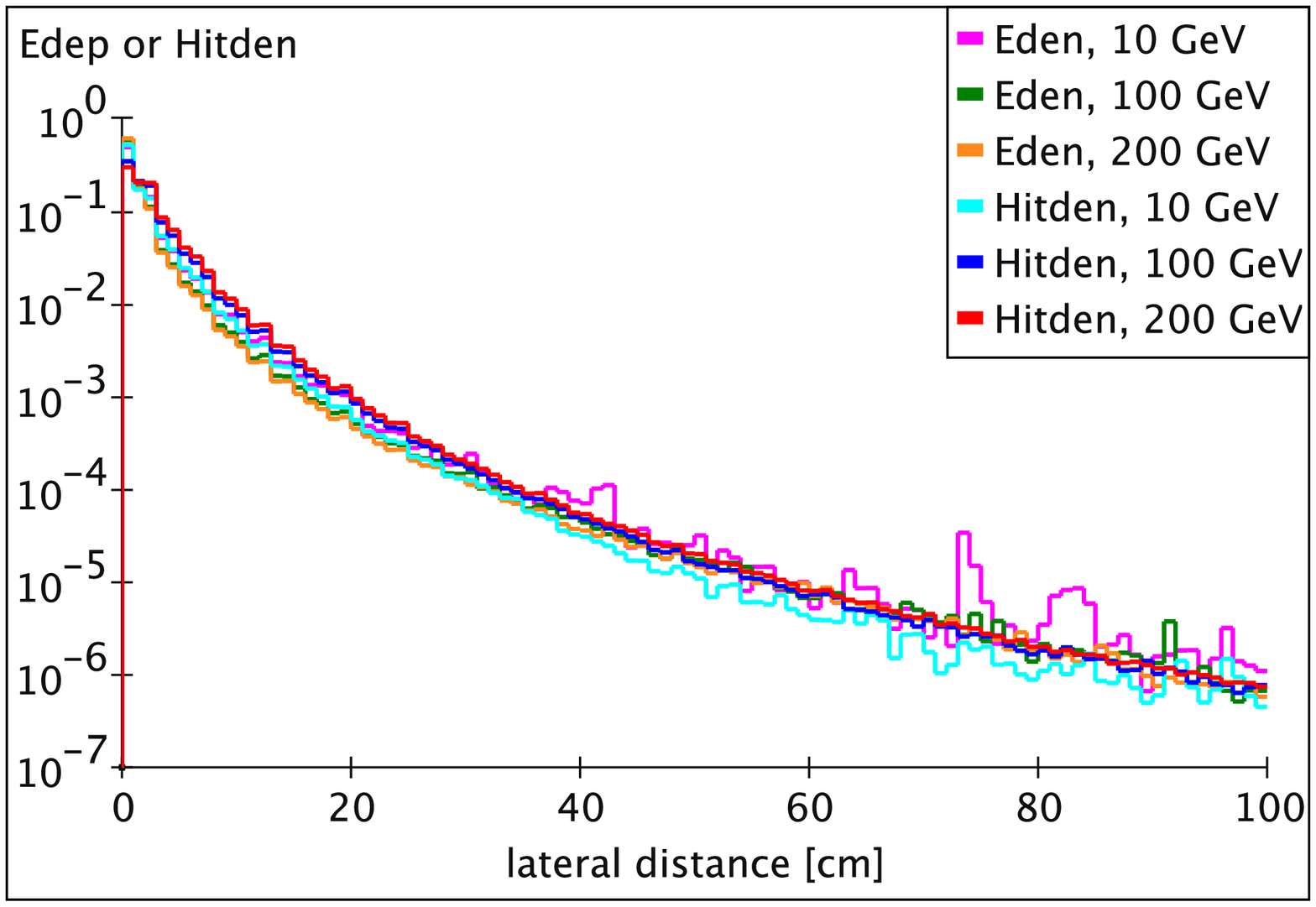}
    \hfill
    \includegraphics[width=0.49\columnwidth]{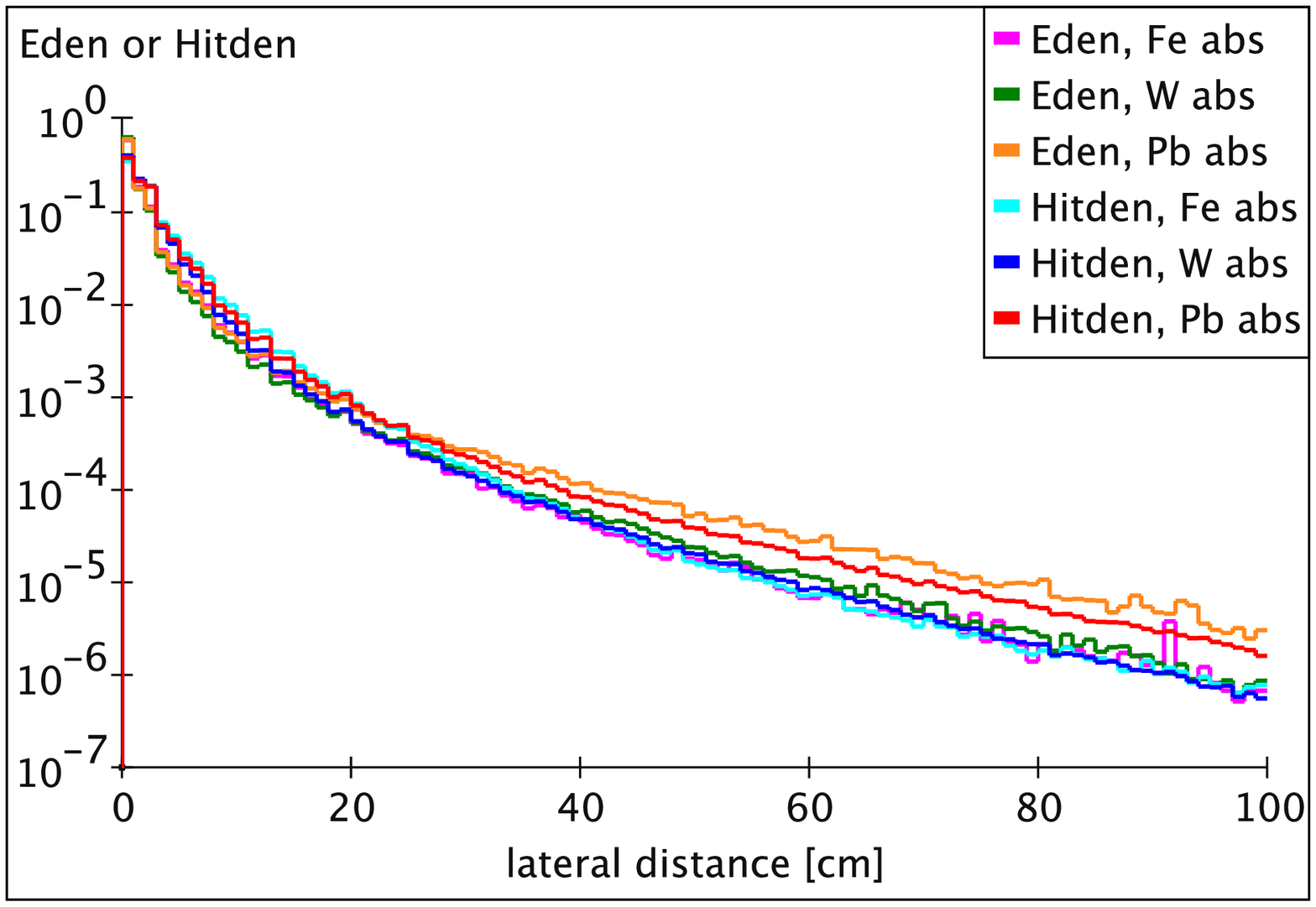}\\ 
    \caption{Longitudinal (top) and lateral (bottom) shower profiles for analog 
      ($E_{dep}$ or $E_{den}$) and digital ($nbHits$ or $Hitden$) readout. The 
      profiles on the left are for Fe absorber and for different pion energies. 
      The profiles on the right are for 100\gev pions and various absorbers.
      The distributions are normalized to 1.}
    \label{profile}
  \end{figure}
\end{center}

The longitudinal shower profile is a sum of deposited energy in 3\mm gas gap 
or number of counted hits for all fired cells in one calorimeter layer versus
calorimeter depth expressed in \lint or number of layers. The results follow 
the expected behavior. First, the shower maximum, due to the progressive 
shower development depending on the energy of the incident particle, is getting 
deeper with increasing energy of primary pions (see Fig.\ref{profile} top 
left). Second, the maximum is deeper for Fe absorber in comparison 
with W and Pb absorbers. This is consequence of the higher $Z$ number and 
smaller \xrad/\lint ratio for W and Pb absorbers in comparison with Fe absorber 
(see Fig.\ref{profile} top right). 

The comparison of analog and digital readout shows that longitudinal shower
profiles are very similar for lower pion energy. For higher energy a shift 
(the shower maximum is deeper for digital with respect to analog readout) for 
higher energy has been observed for all three absorbers (see Fig.\ref{profile} 
top). The shift, which is probably due to the saturation, is getting to be more 
important with increasing pion energy.

The lateral shower profile, described as the energy or hit density (i. e. 
deposited energy or number of counted hits per unit area) as a function of 
radial distance from beam axis, shows an expected narrow core where mainly the 
electromagnetic component of the hadronic shower contributes. The core 
is surrounded by a gradually decreasing halo for which the hadronic component 
is responsible (see Fig.\ref{profile} bottom). Similar behavior has been 
found for Fe and W absorbers in comparison with Pb showing slightly higher 
density for larger lateral distance. Also a shift between analog and 
digital readout has been found for lateral profile and can be explained by 
the same reason as in case of the longitudinal shower profile.

\subsection{Longitudinal fractional deposited energy}
\begin{center}         
  \begin{figure}[htb]
    \includegraphics[width=0.49\columnwidth]{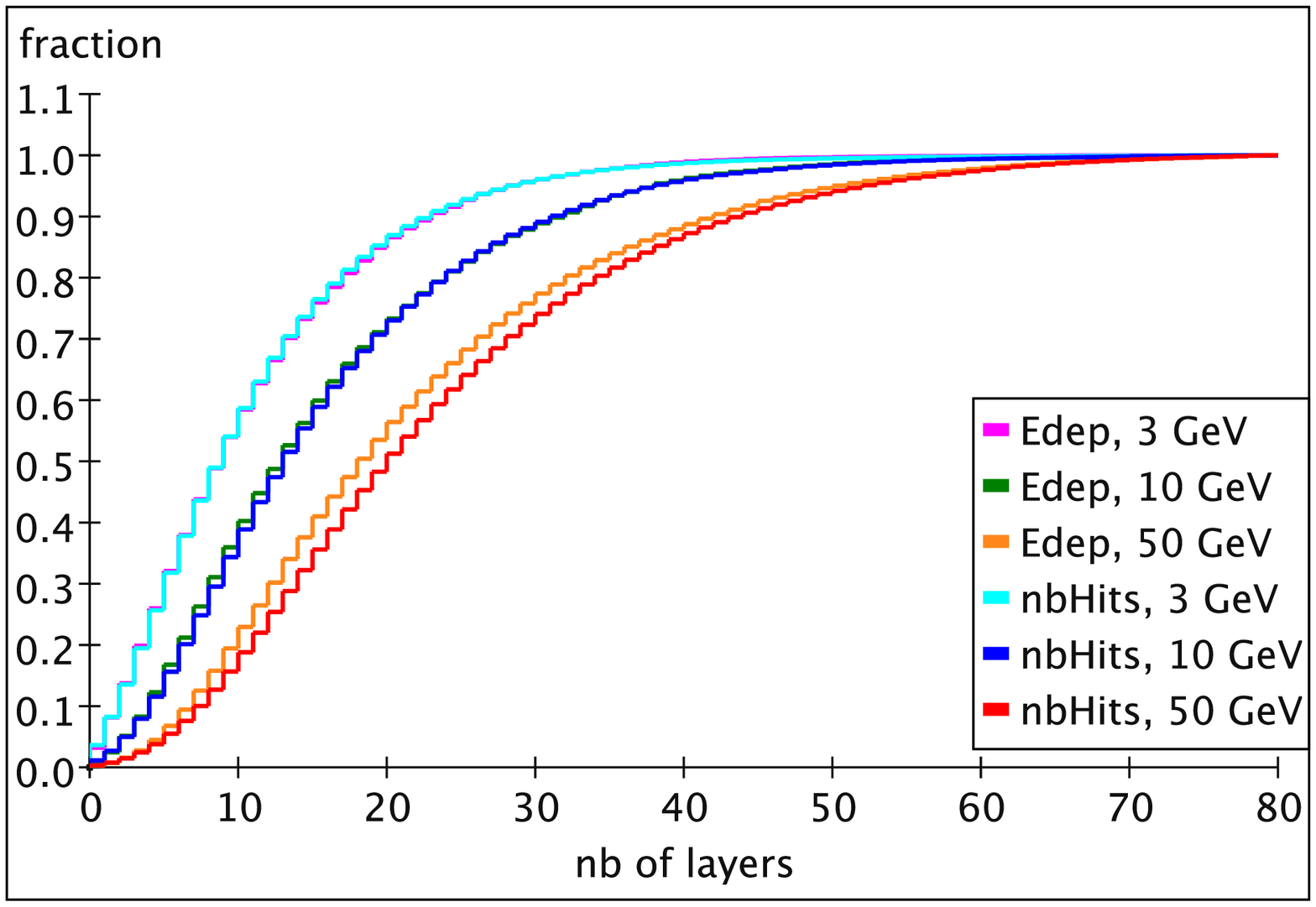}
    \hfill
    \includegraphics[width=0.49\columnwidth]{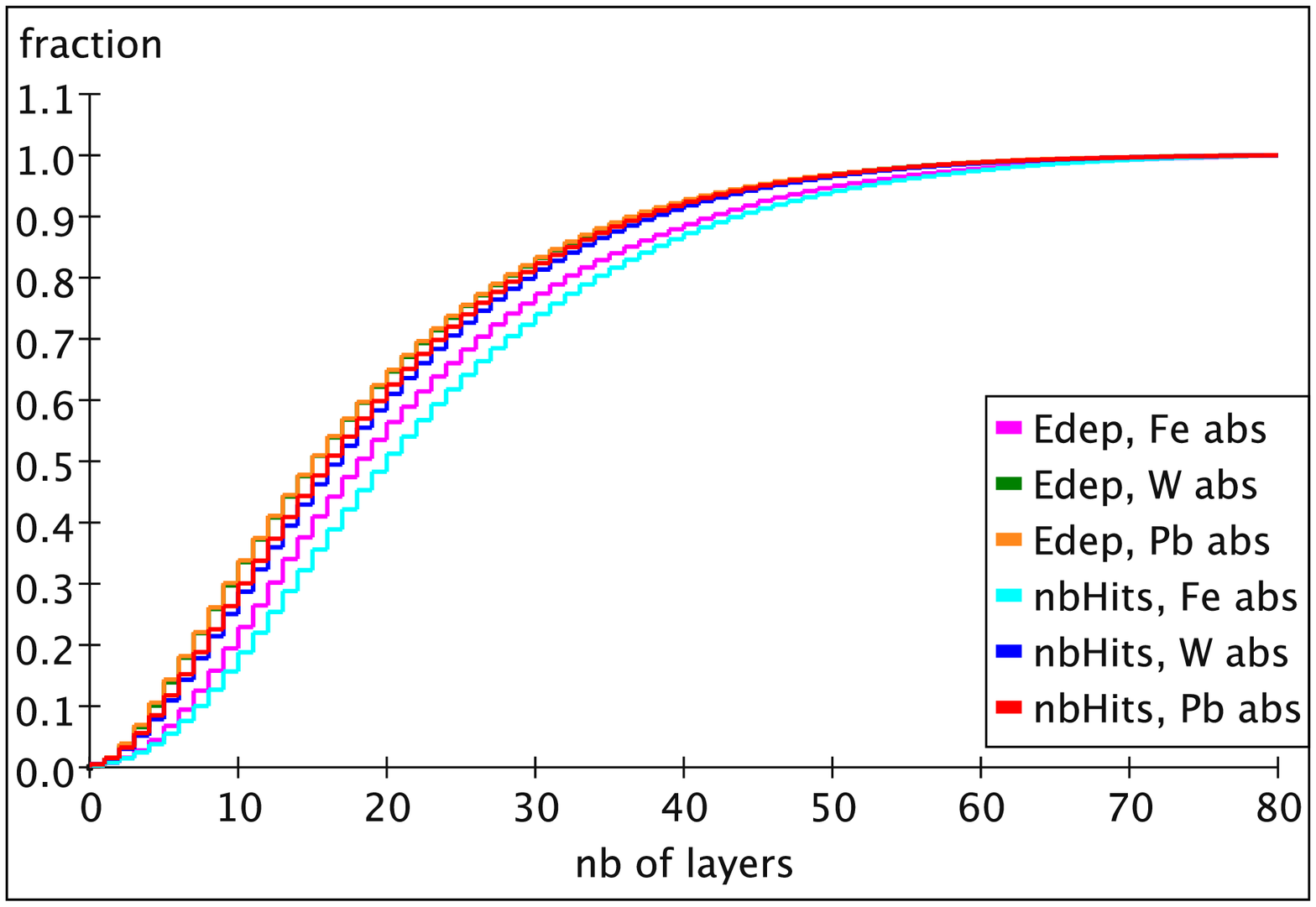}\\ 
    \caption{Longitudinal fractional deposited energy in calorimeter for analog 
      ($E_{dep}$) and digital ($nbHits$) readout for Fe absorber and different 
      pion energies (left), and for various absorbers and 50\gev pions (right).}
    \label{fraction}
  \end{figure}
\end{center}

The longitudinal fractional deposited energy shows the fraction for a 
calorimeter of chosen depth (in \lint or number of layers) with respect 
to the maximal calorimeter depth (9\lint or 80 layers). For primary pions, 
which deposit almost 100\% of their energy in 9\lint, the fractional 
deposited energy can be approximately equal to the energy containment. If 
this is assumed to be true for 50\gev pions, the 95\% energy containment 
can be reached with calorimeter having 50 layers (about 5.6\lint) equipped 
with Fe absorber or 45 layers (about 5\lint) in case of W or Pb absorbers 
(see Fig.\ref{fraction}). In Fig.\ref{fraction}, the shift between analog and 
digital readout is also seen, which is linked to the effect discussed above. 
This shift can lead to an underestimation of the calorimeter depth if only 
information from digital readout is considered. 

\section{Energy resolution and linearity}
\subsection{Deposited energy in analog and digital mode}
\begin{center}              
  \begin{figure}[htb]
      \includegraphics[width=0.33\columnwidth]{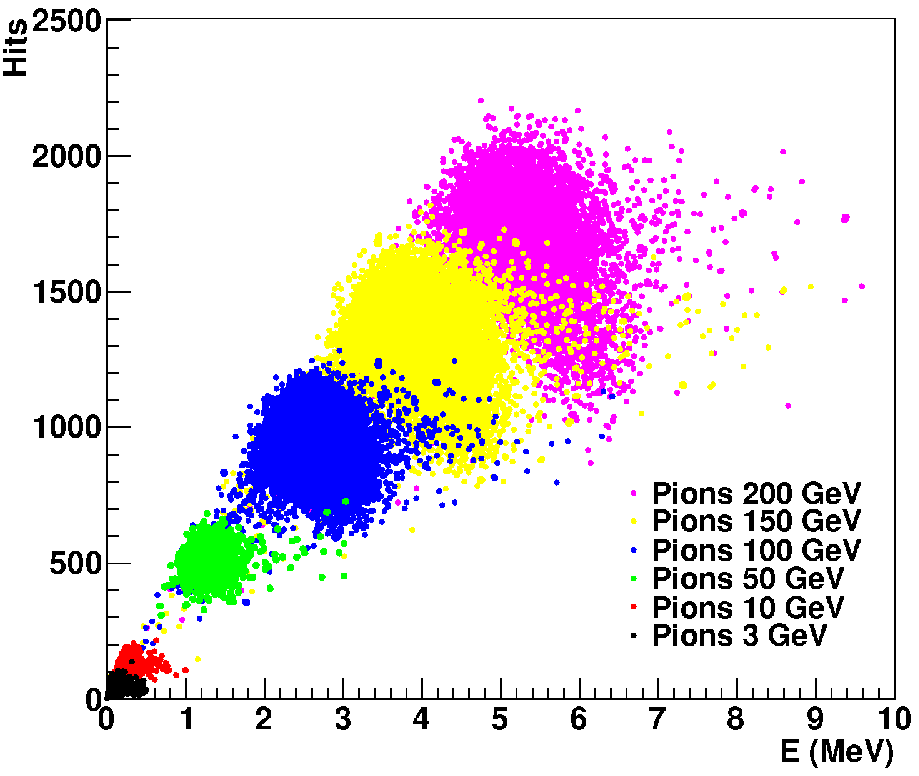}
      \hfill
      \includegraphics[width=0.33\columnwidth]{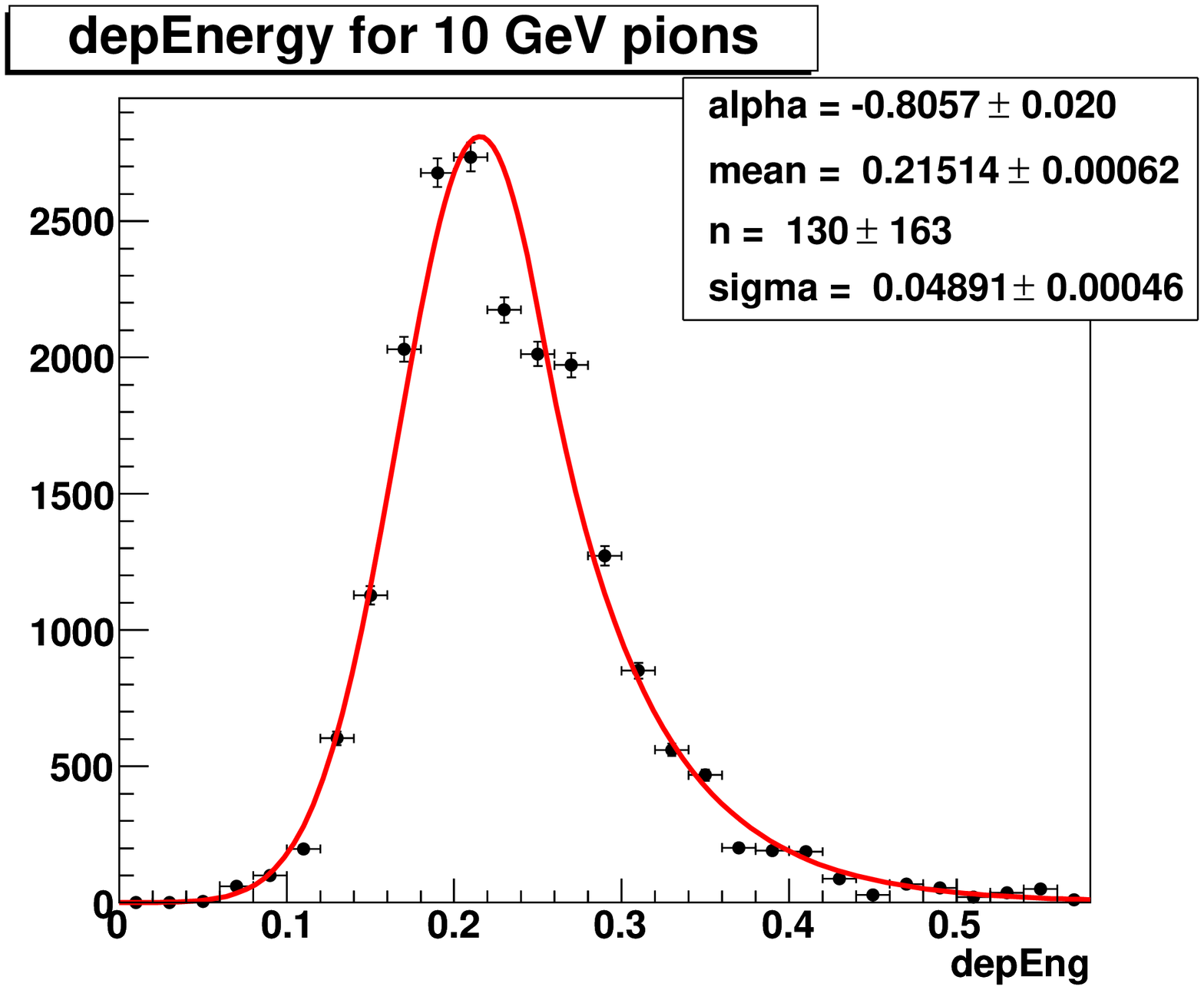} 
      \hfill
      \includegraphics[width=0.33\columnwidth]{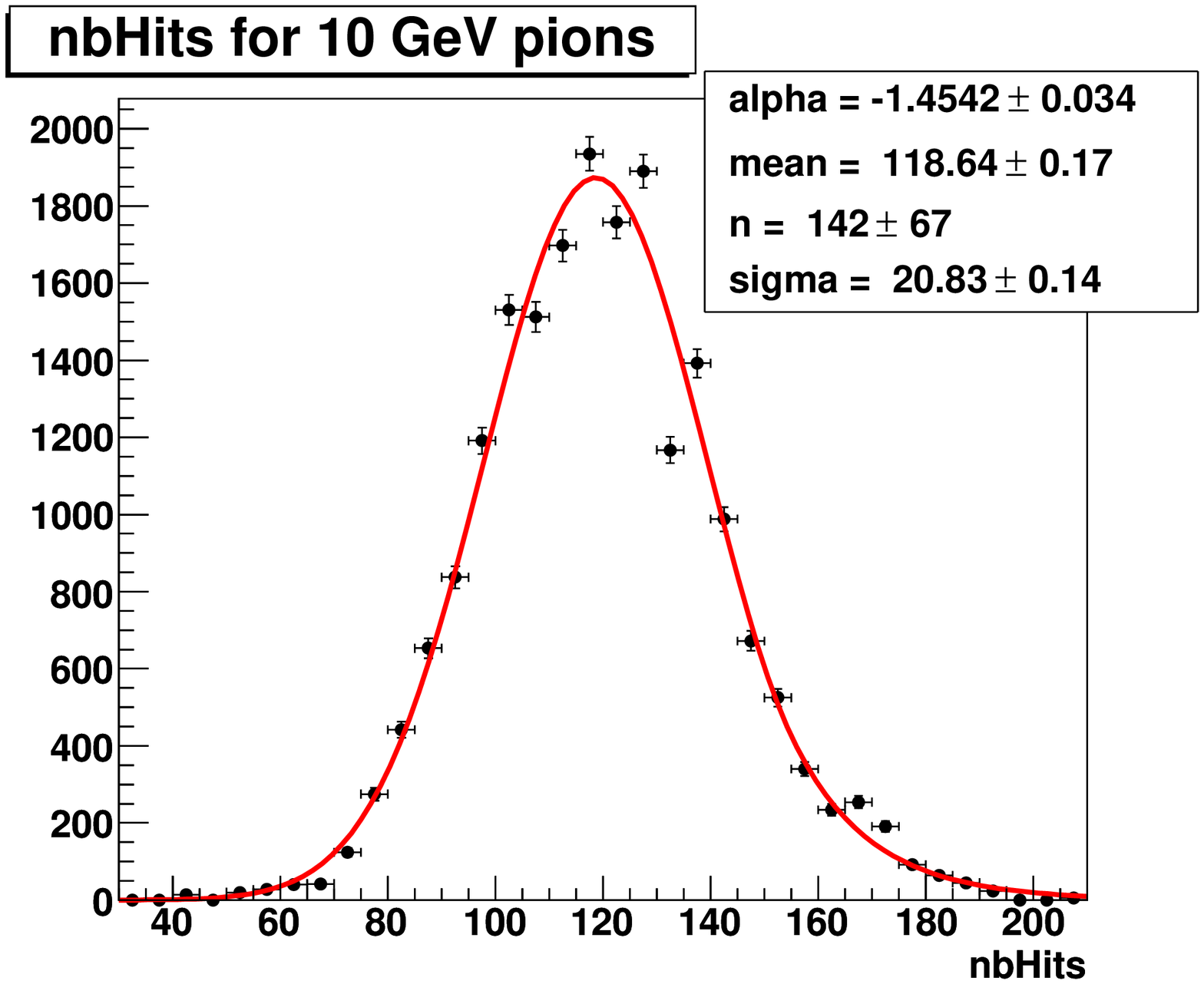}
      \\
      \caption{Left: Correlation between deposited energy in Fe calorimeter and 
        number of counted hits for pions energies. Middle and right: 
        Distributions of deposited energy (analog readout) and number of 
        counted hits (digital readout) in calorimeter with Fe absorber for 
        10\gev pions.}
      \label{distr}
  \end{figure}
\end{center}

The energy measured in a digital calorimeter, where only cells with energy
above a chosen threshold are counted, is based on the very simple idea 
that the number of hits (fired cells) is directly proportional to the energy 
deposited in active medium and thus to the total energy absorbed in the 
calorimeter. The correlation between energy deposited in active medium 
(3\mm of gas) and number of counted hits for calorimeter with Fe absorber 
and for pion energy from 3 to 200\gev is shown in Fig.~\ref{distr} left.

A distribution of the energy measured in hadronic calorimeter with gaseous 
detector presents significant right-hand tail due to the large Landau 
fluctuations in energy deposition in gas (see Fig.~\ref{distr} middle). 
The digital readout leads to the suppression of these fluctuations, and 
consequently to an improvement of the energy resolution (see Fig.~\ref{distr} 
right). On the other hand, the energy resolution at higher energies in 
digital mode is affected by saturation of the number of counted hits 
(see Fig.~\ref{response} right). 

\subsection{Response and linearity}
\begin{center}              
  \begin{figure}[tbh]
      \includegraphics[width=0.5\columnwidth]{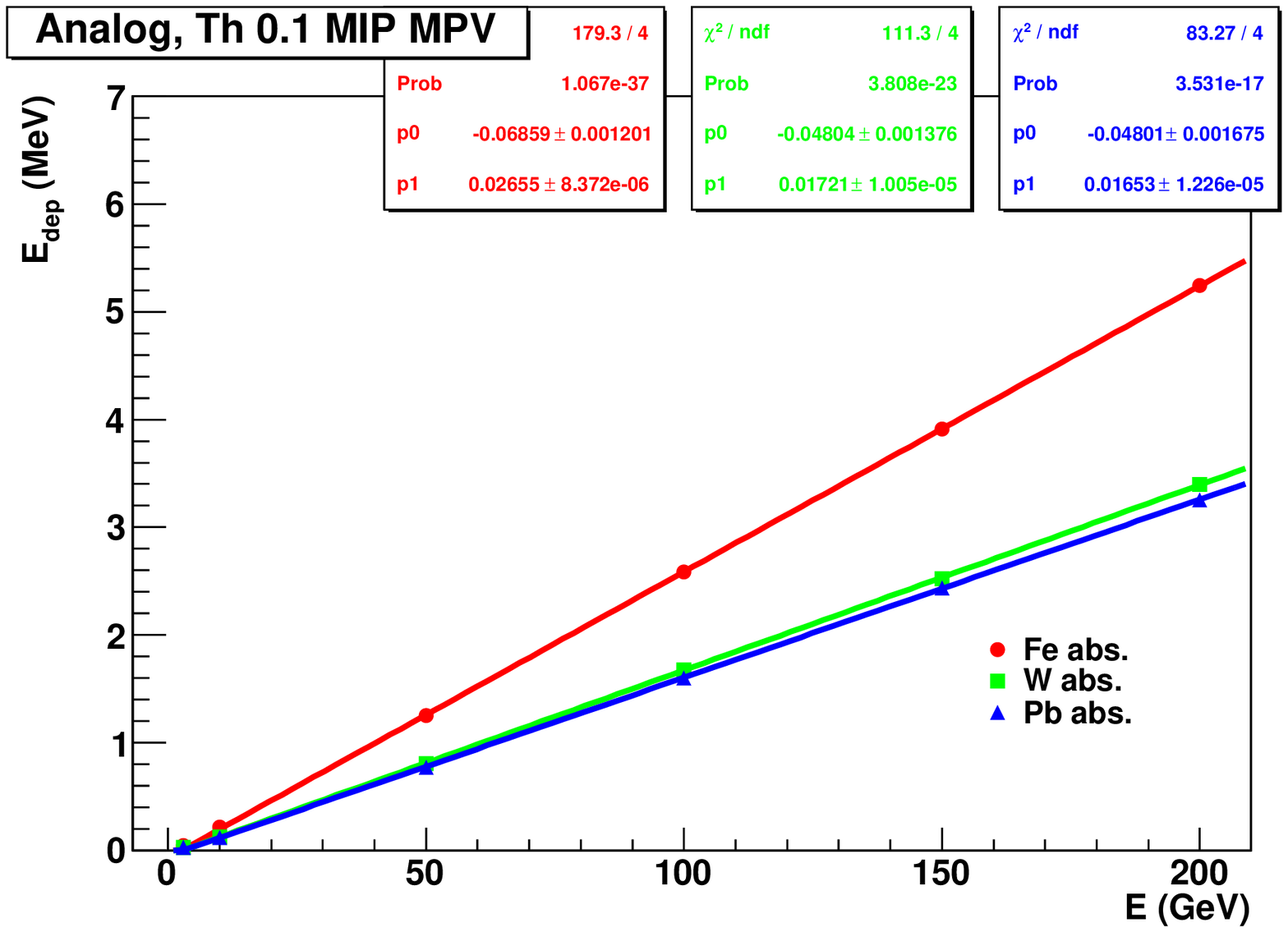} \hfill
      \includegraphics[width=0.5\columnwidth]{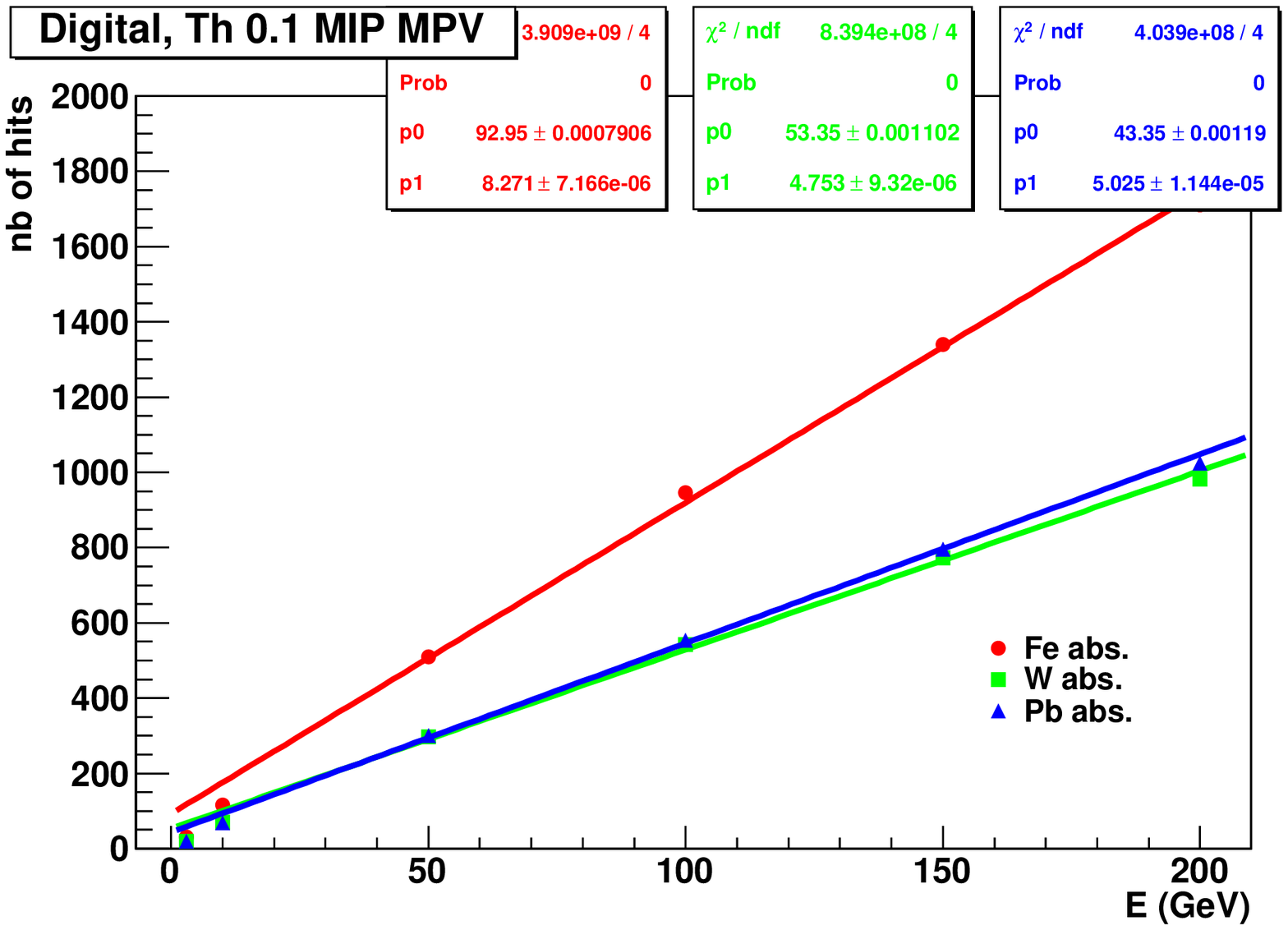} \\
      \includegraphics[width=0.5\columnwidth]{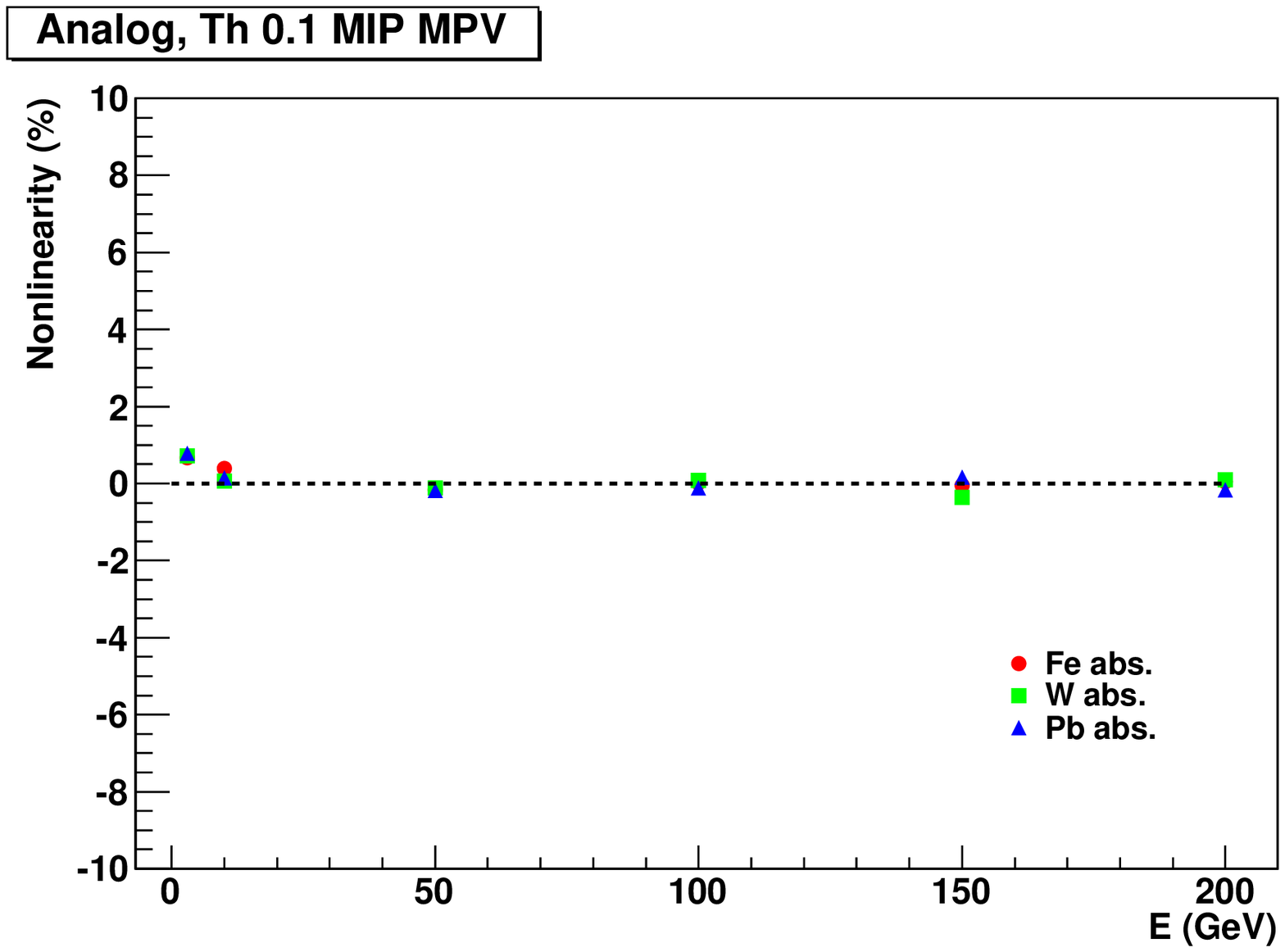} \hfill
      \includegraphics[width=0.5\columnwidth]{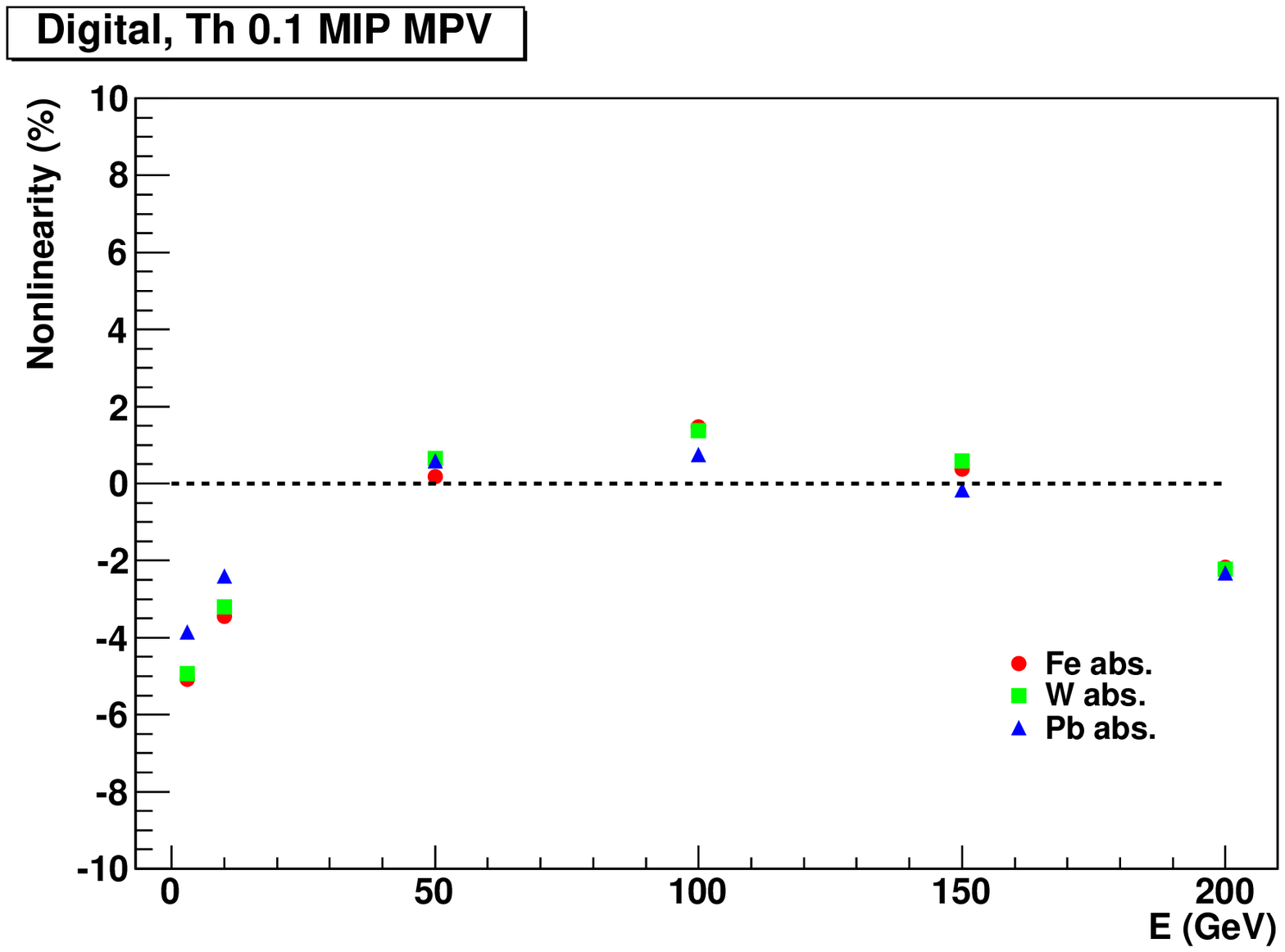} \\
      \caption{Calorimeter response (top) and full scale non-linearity (bottom) 
        for analog (left) and digital (right) readout and for various absorber
        materials.}
      \label{response}
  \end{figure}
\end{center}

The linear relation between calorimeter response and energy of primary pions 
for analog and digital readouts is shown in Fig.~\ref{response} top. The amount
of energy deposited or number of hits is significantly higher in case
of Fe absorber which is due to its longer \xrad. The higher number of 
hits in case of Fe absorber and the properties of this material (\xrad, \lint) 
have a positive impact on the energy resolution described in following section. 

The linearity of the response was quantified by the full-scale non-linearity, 
i.\,e. the residuals of the linear fit of the response vs pion energy divided 
by the response of the higher primary pion energy (200\gev). As can be seen 
in Fig.~\ref{response} bottom, the non-linearity behaves similar for all the
absorbers and is within $\pm$1\% for analog and $\pm$5\% for digital readout, 
respectively. The worse linearity for digital readout can be due to a 
saturation effect, when the number of hits does not follow increasing
energy of the incident particles. This could be improved by adding one or 
two more thresholds ({\it semi-digital} readout).

\subsection{Energy resolution}
\begin{center}              
  \begin{figure}[tbh]
      \includegraphics[width=0.49\columnwidth]{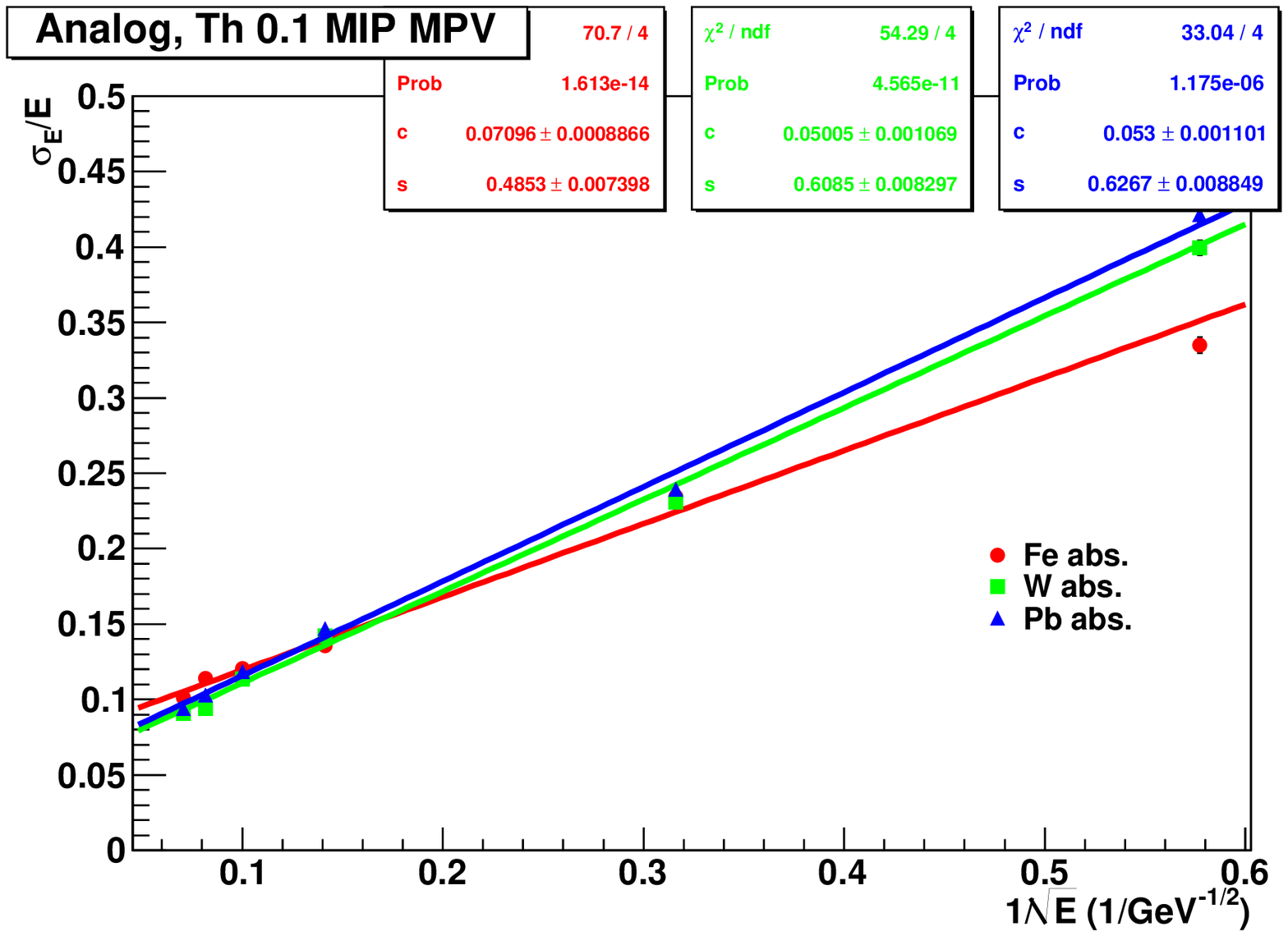} \hfill
      \includegraphics[width=0.49\columnwidth]{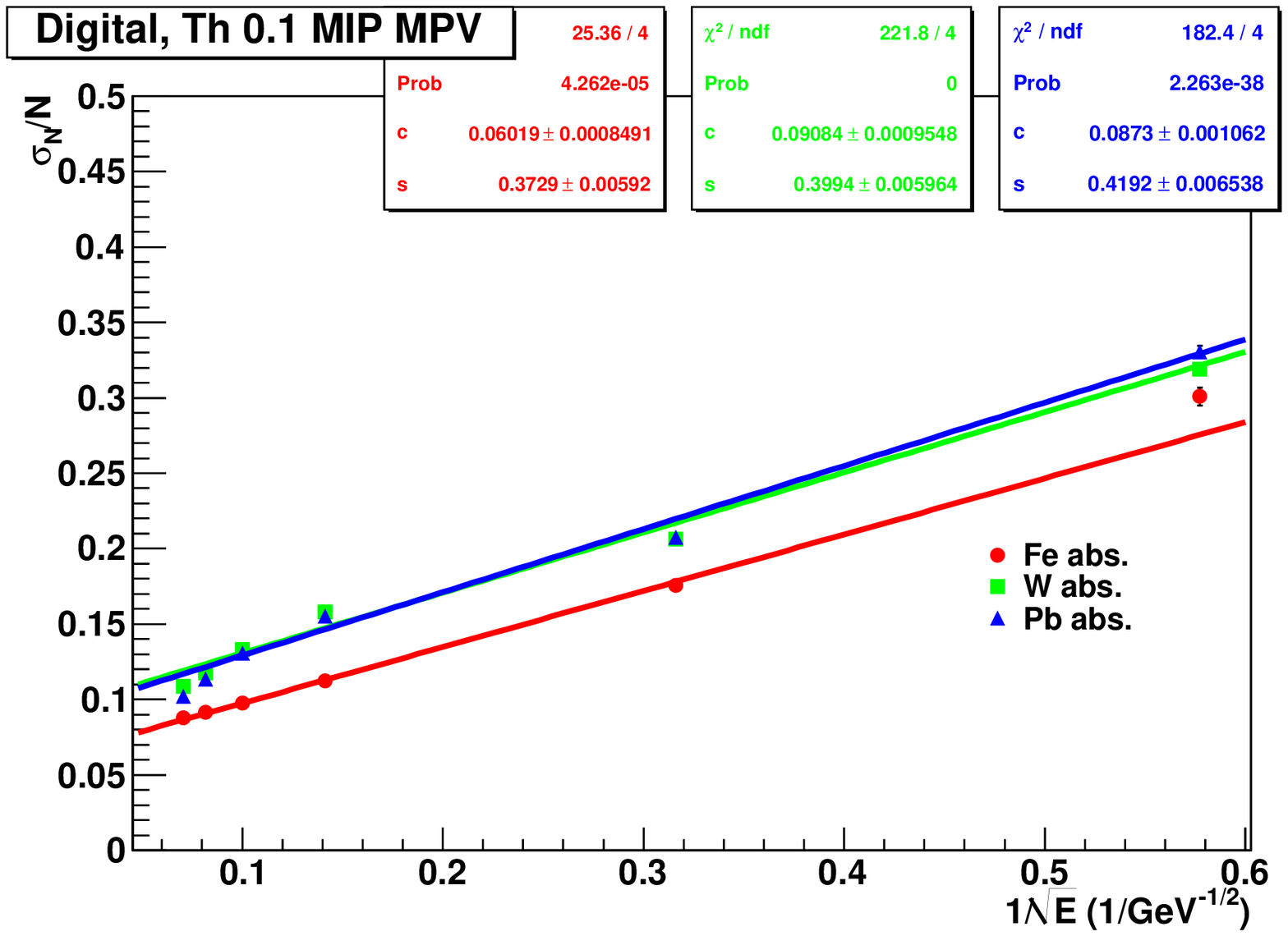} \\
      \caption{Energy resolution as a function of reciprocal square root of 
        pion energy for analog (left) and digital (digital) readout and for 
        various absorber materials.}
      \label{resolution}
  \end{figure}
\end{center}

The energy resolution as a function of pion energy for various absorber 
materials and different readouts is displayed in Fig.\ref{resolution}. 
The standard parametrization, $ \sigma_{E}/E = S/\sqrt{E} \oplus N/E \oplus C$,
for energy resolution as a function of incident particle is used in order to 
extract and evaluate the stochastic $S$ and constant $C$ terms 
(numeric values of these parameters are shown in Fig.\ref{resolution}).
Since any kind of noise is not present in the simulation, the noise term $N$ 
in parametrization is not taken into account.

In case of analog readout, the energy resolution is similar for W and Pb 
absorbers and slightly different for Fe absorber as is shown in 
Fig.\ref{resolution} left. An identical behavior has been found for W and Pb 
absorbers with digital readout (see Fig.\ref{resolution} right). Comparing 
analog and digital readout for these absorbers, it has been observed that 
analog readout performs always better at high energy. This is a consequence 
of the suppression of Landau fluctuations and the saturation, respectively. 
On the other hand, the energy resolution for Fe absorber with digital readout 
is superior over the whole energy range as a consequence of the higher number 
of counted hits due to the longer \xrad and larger $R_M$ of Fe with respect to 
W and Pb.

\section{Summary and conclusions}
The topology of the hadronic shower can be well described also in digital mode.
Small difference in shower profiles between analog and digital mode have been 
found. The energy resolution for digital in comparison with analog readout 
tends to be superior for lower and inferior for higher energy. The linearity 
has been found always better for analog in comparison with digital readout.
 
Generally, it can be concluded that the presented Monte Carlo study has proved 
that a DHCAL concept, with respect to the basic performance characteristics, 
fulfills linear collider detector requirements. A difference in performance 
between digital and analog approaches has been identified and will be a 
subject of further investigation.


\end{document}